\documentclass[submission,copyright,creativecommons]{eptcs}
\providecommand{\event}{ICLP 2025} % Name of the event you are submitting to
\usepackage{tikz}
\usepackage{listings}
\usepackage{fancyvrb}
\usepackage[utf8]{inputenc}
\usepackage{orcidlink}
\usepackage{iftex}
\usepackage{url} 
\ifpdf
  \usepackage{underscore}         % Only needed if you use pdflatex.
  \usepackage[T1]{fontenc}        % Recommended with pdflatex
\else
  \usepackage{breakurl}           % Not needed if you use pdflatex only.
\fi

\title{Picat Through the Lens of Advent of Code}
\author{Neng-Fa Zhou \orcidlink{0000-0003-2507-7031}
\institute{CUNY Brooklyn College and Graduate Center}
\email{nzhou@acm.org}
\and
Cristian Grozea \orcidlink{0000-0001-6393-1919}\thanks{Cristian Grozea's work has been partially funded by the German Federal Ministry of Education and Research (BMBF, Germany) as part of the 6G Research and Innovation Cluster 6G-RIC under Grant 16KISK020K.} 
\institute{Fraunhofer Institute FOKUS\\
Berlin, Germany
}
\email{cristian.grozea@fokus.fraunhofer.de}
\and
H{\aa}kan Kjellerstrand \orcidlink{0000-0003-1804-7117}
\institute{Independent Researcher, hakank.org}
\email{hkjellerstrand@acm.org}
\and
Oisín Mac Fhearaí
\email{denpashogai@gmail.com}
}

\newcommand{\titlerunning}{Solving AoC with Picat}
\newcommand{\authorrunning}{Zhou, Grozea, Kjellerstrand, Mac Fhearaí}

\hypersetup{
  bookmarksnumbered,
  pdftitle    = {\titlerunning},
  pdfauthor   = {\authorrunning},
  pdfsubject  = {EPTCS},               % Consider adding a more appropriate subject or description
}

\begin{document}
\maketitle

\begin{abstract}
Picat is a logic-based, multi-paradigm programming language that integrates features from logic, functional, constraint, and imperative programming paradigms. This paper presents solutions to several problems from the 2024 Advent of Code (AoC). While AoC problems are not designed for any specific programming language, certain problem types, such as reverse engineering and path-finding, are particularly well-suited to Picat due to its built-in constraint solving, pattern matching, backtracking, and dynamic programming with tabling. This paper demonstrates that Picat's features, especially its SAT-based constraint solving and tabling, enable concise, declarative, and highly efficient implementations of problems that would require significantly more effort in imperative languages.
\end{abstract}

\section{Introduction}
The Advent of Code (AoC) is an annual programming challenge that presents a series of algorithmic problems\cite{adventofcode}, attracting programmers from diverse backgrounds worldwide. The problems range from simple puzzles to complex computational challenges, making them an excellent benchmark for evaluating programming languages and problem-solving paradigms. A defining feature of AoC is that each daily challenge has a first part, that is usually easier to solve, and a second part, that comes after a twist of the original problem proposed in the first part, making it generally much more difficult to solve and requiring flexibility such that the solution or at least the model defined for the first part can be reused.

Picat \cite{PicatWebsite}, a logic-based multi-paradigm programming language, offers a unique combination of logic, functional, constraint, and imperative programming styles. It is particularly well-suited for solving AoC problems due to its built-in constraint solving, pattern matching, backtracking, and dynamic programming with tabling features. These capabilities allow for concise, declarative, and highly efficient implementations of problems that would require significantly more effort in imperative languages.

In this paper, we explore Picat solutions to five selected problems from the 2024 Advent of Code, demonstrating how Picat’s distinctive features provide elegant and efficient solutions. While there is work on the usage of Picat for various kinds of problems \cite{DymchenkoM15,Zhou16}, we demonstrate in this paper some of the new features of Picat added over the past decade. Specifically, we apply Picat’s SAT-based constraint solver to solve three combinatorial problems, showcasing its ability to efficiently model and solve constraint satisfaction problems. Additionally, we leverage tabling to solve two problems that require dynamic programming, illustrating how Picat’s built-in tabling mechanism simplifies recursive computations and optimizes performance.

This paper presents Picat solutions for the following problems: \textbf{Day 23: LAN party}, a clique-finding graph problem; \textbf{Day 17: Chronospatial Computer}, a reverse-engineering puzzle; \textbf{Day 24: Crossed Wires}, a circuit simulation and repair problem; \textbf{Day 16: Reindeer Maze}, a path-finding problem; \textbf{Day 21: Keypad Conundrum}, an optimal planning problem. Through these case studies, we aim to highlight the expressiveness and power of Picat’s constraint solving and tabling features, demonstrating its advantages in tackling complex algorithmic challenges efficiently. For each problem, we also give a brief comparison with other solutions in terms of efficiency and needed efforts. All the CPU times reported in this paper were measured on an Ubuntu Linux machine with an Intel i9-7940X CPU @ 3.10GHz and 64Gb RAM. For SAT-based solutions, Kissat\footnote{\url{https://github.com/arminbiere/kissat}} is used.

\section{The Picat Language}
This section gives an overview of the features of the Picat language. Readers are referred to \cite{PicatGuide} for a detailed description of the language, and \cite{PicatBook15} for uses of Picat in modeling and solving combinatorial problems.

Picat is a logic-based multi-paradigm programming language designed for both general-purpose scripting and combinatorial problem modeling.  It is grounded in foundational logic programming technologies, including the design of Prolog \cite{ColmerauerR93,Kowalski79}, the Warren Abstract Machine (WAM) \cite{Warren83,wam:hassan}, Constraint Logic Programming over Finite Domains (CLP(FD)) \cite{HEN89}, and tabling \cite{Warren92}.

Picat supports a rich set of data types, many of which go beyond those traditionally found in Prolog. While it includes standard types such as atoms, numbers, variables, and compound terms (shared with Prolog), Picat introduces several new data types that enhance usability and expressiveness, including mutable arrays, maps (aka dictionaries), and sets.

In Picat, pattern-matching rules are used to define predicates and functions. Picat has two types of pattern-matching rules: the \emph{non-backtrackable} rule\index{non-backtrackable rule} 
\begin{tabbing}
aa \= aaa \= aaa \= aaa \= aaa \= aaa \= aaa \kill
\> \> $Head, Cond\ $\verb+=>+$\ Body$. 
\end{tabbing}
and the \emph{backtrackable} rule\index{backtrackable rule} 
\begin{tabbing}
aa \= aaa \= aaa \= aaa \= aaa \= aaa \= aaa \kill
\> \> $Head, Cond\ $\verb+?=>+$\ Body$. 
\end{tabbing}
For a call, a rule is applicable if the pattern in $Head$ matches the call and the condition $Cond$ is true. Once an applicable rule is found, Picat executes $Body$. If the selected rule is non-backtrackable, Picat commits to this rule, meaning that no other rules will be considered, even if $Body$ fails. If the selected rule is backtrackable, Picat backtracks when $Body$ fails, and continues to search for the next applicable rule. Picat also supports Prolog-style Horn clauses.

Picat distinguishes between functions, which return values, and predicates, which define relations and may have multiple solutions. A predicate head has the form $p(T_1, \ldots, T_n)$, where $p$ is a predicate symbol and each $T_i$ is an argument pattern. A function head has the form $f(T_1, \ldots, T_n) = R$, where $f$ is a function symbol, each $T_i$ is an argument pattern, and $R$ is the return value. If both the condition $Cond$ and the body $Body$ are true, the rule can be abbreviated as $f(T_1, \ldots, T_n) = R$. Unlike Prolog, where a structural term automatically constructs a compound term, Picat uses the \verb+$+ symbol before a structural term to explicitly denote structure construction. So while \texttt{max(X,Y)} is a function call, \verb+$max(X,Y)+ is a structure construction.

Picat supports list and array comprehensions, enabling concise construction of complex data structures through declarative expressions. For example, the following array comprehension reads a maze map as input and converts it to a two-dimensional array:
\begin{verbatim}
    M = {to_array(L) : L in read_file_lines(File)}
\end{verbatim}

Picat provides loops to support imperative-style programming alongside its declarative features. A \texttt{foreach} loop has the following general form:
\begin{tabbing}
aaaa \= aaa \= aaa \= aaa \= aaa \= aaa \= aaa \kill
\> \texttt{foreach ($E_1$ in $D_1$, $Cond_1$, $\ldots$, $E_n$ in $D_n$, $Cond_n$)}  \\
\> \> $Goal$ \\
\> \texttt{end} 
\end{tabbing}
The expression \texttt{$E_i$ in $D_i$} is called an \emph{iterator}, where $E_i$ is an iterating pattern, and $D_i$ is an expression that gives a compound value.  Each $Cond_i$ is an optional condition on iterators $E_1$ through $E_i$. It is used for looping through collections like lists, arrays, and ranges. Picat compiles loops away by transforming them into new tail-recursive predicate definitions \cite{ZhouF17}.

Picat supports the assignment operator, which is particularly useful in loops where a variable’s value needs to be updated over time. An assignment in the form $LHS := Exp$ assigns the value of $Exp$ to $LHS$. If $LHS$ is an indexing operation of the form $X[I_1, \ldots, I_n]$, Picat updates the value stored at the specified location within the indexed structure. If $LHS$ is a variable, Picat introduces a new variable to hold the value of $Exp$ and replaces all subsequent occurrences of the original variable with this new one. In this latter case, where $LHS$ is a variable, Picat compiles the assignment away by transforming it into new predicate definitions \cite{ZhouF17}.

These new data types, combined with Picat’s support for pattern matching, comprehensions, and imperative constructs, make the language more expressive and practical for a wide range of programming tasks, especially those involving structured data and numeric computation. 

Picat provides a unified interface for CP, SAT, MIP, and SMT solvers, allowing seamless solver switching for constraint satisfaction problems. It includes an award-winning SAT-based CSP solver, PicatSAT\cite{zhou2016picat,ZhouK17}, with efficient encodings for high-level constraints.

Picat offers a tabling mechanism for dynamic programming solutions. It incorporates structure sharing for tabled data to efficiently manage memory and avoid state explosion \cite{Zhou12}. The tabling-based planner of Picat provides a number of strategies for efficiently solving planning problems using state-space search \cite{ZhouBD15}.

These features make Picat a powerful tool for both general-purpose programming and combinatorial problem modeling.

\section{Day 23: LAN Party}
In Advent of Code 2024, Day 23, the problem revolves around analyzing a network of computers connected by bidirectional links. The input describes these connections, forming an undirected graph, and the goal is to extract meaningful structural properties from this network.

Part-1 is to identify all unique triads—sets of three computers that are all directly connected to one another. Additionally, the problem requires counting how many of these triads include at least one computer whose name starts with 't'.  This part can be trivially solved in any programming language by counting triads involving each of the vertices.

Part-2 escalates the challenge by requiring the identification of a maximum clique in the network. A clique is a subset of nodes in which every pair of nodes is directly connected, and the maximum clique is the largest such subset. Finding a maximum clique is an NP-hard problem, making it a suitable challenge for constraint programming and graph algorithms.

Our Picat solution is given below.
%\footnote{The full source code is available at \url{https://github.com/nfzhou/aoc/blob/main/aoc_24_23_part2.pi}}

\noindent
\rule{\textwidth}{1pt}
\begin{small}
\begin{verbatim}
% A Picat program for the maximum clique problem.

import sat.

main([File]) =>
    input(File,VSet,EdgeSet),
    maxClique(VSet,EdgeSet,Bs),
    output(VSet,EdgeSet,Bs).

maxClique(VSet,EdgeSet) =>
    Vs = to_array(keys(VSet)),
    N = len(Vs),
    Bs = new_array(N),
    Bs :: 0..1,
    foreach (I in 1..N-1, J in I+1..N)
        if not EdgeSet.has_key([Vs[I],Vs[J]]) then
            #~Bs[I] #\/ #~Bs[J]
        end
    end,
    Count #= sum([B : B in Bs]),
    solve([$max(Count)],Bs).
\end{verbatim}
\end{small}
\rule{\textwidth}{1pt}

\noindent
The first line imports the \texttt{sat} module. Picat supports four solver modules: \texttt{sat}, \texttt{cp}, \texttt{mit}, and \texttt{smt}. The import statement specifies that the SAT-based solver should be used for this program.

The \texttt{input(File,VSet,EdgeSet)} predicate reads a graph specification from \texttt{File} into \texttt{VSet} and \texttt{EdgeSet}, where \texttt{VSet} is a hash set of vertices, and \texttt{EdgeSet} is a hash set of edges.

The \texttt{maxClique(VSet,EdgeSet)} predicate formulates the maximum clique problem as a constraint optimization problem. It assigns a binary variable to each vertex, where 1 indicates the vertex is included in the clique and 0 otherwise. For any two vertices \texttt{I} and \texttt{J} that are not connected by an edge, let \texttt{Bs[I]} and \texttt{Bs[J]} be their corresponding binary variables. The model enforces that these two unconnected vertices cannot both be included in the clique using the constraint: \verb+#~Bs[I] #\/ #~Bs[J]+. The objective is to maximize the sum of the binary variables.

It took the Picat program 10.8 seconds to find a maximum clique in the input graph, which is made of 520 vertices and 3380 edges. When Kissat is switched to a max-sat using CGSS2\footnote{To switch to maxsat, change \texttt{solve([\$max(Count)],Bs)} to \texttt{solve([\$max(Count),maxsat],Bs)}.}, the program took 2.01 seconds. Some of the solutions in other languages on Reddit\footnote{\url{https://www.reddit.com/r/adventofcode/comments/1hkgj5b/2024_day_23_solutions/}}  are much faster than our SAT-based solution. For example, the solution based on the Bron-Kerbosch algorithm, a recursive backtracking algorithm, only took 0.055 seconds. Even the brute force solution\footnote{\url{https://github.com/tmo1/adventofcode/blob/main/2024/23b.py}}, which recursively expands cliques, took the same amount of time. While our solution is not competitive on the input graph in terms of speed, it is a straightforward declarative solution that can be written without knowledge of specialized graph algorithms.

\section{Day 17: Chronospatial Computer}
In Advent of Code 2024, Day 17, participants are tasked with analyzing a program designed for an imaginary machine called a Chronospatial Computer (CC). This CC computer operates with three registers A, B, and C, which can each hold any integer value. Its program consists of a sequence of 3-bit numbers (ranging from 0 to 7), each representing either an instruction opcode or an operand. For example, the following is a CC program, where the left column gives instructions and the right column gives the representing symbols.

\noindent
\begin{center}
\begin{tabular}{|l|l|} \hline
2,4 &   \texttt{B = A mod 8} \\
1,5 &   \texttt{B = B\string^ 5} \\
7,5 &   \texttt{C = A/2**B} \\
0,3 &   \texttt{A = A/2**3} \\
4,0 &   \texttt{B = B \string^ C} \\
1,6 &   \texttt{B = B \string^ 6} \\
5,5 &   \texttt{out B mod 8} \\
3,0 &   \texttt{jnz 0} \\ \hline
\end{tabular}
\end{center}
\noindent 

\noindent
The last instruction \texttt{"jnz 0"} sets the instruction pointer to the first instruction if register A contains a non-zero value.

The challenge is divided into two parts, focusing on different aspects of program execution and reverse engineering. Part-1 requires careful implementation of the instruction set to accurately trace the program's behavior and determine the resulting output sequence. Part-2 presents a reverse engineering challenge. Given a specific output sequence, participants must deduce the smallest possible initial value of register A that, when the program is executed, produces the program itself as the output. This involves understanding the relationship between the initial state and the program's output, effectively working backward from the output to infer the input. As the output is the same as the program itself, the sequence of instructions up to \texttt{"out B mod 8"} needs to be executed 16 times before register A holds the value 0 and the program terminates.

Our Picat solution for the above CC program is given below.

\noindent
\rule{\textwidth}{1pt}
\begin{small}
\begin{verbatim}
% A Picat program for Chronospatial Computer reverse engineering.

import sat.

main =>
    A = new_bv(56),
    gen(A,[2,4,1,5,7,5,0,3,4,0,1,6,5,5,3,0]),
    solve($[min(A)],A),
    println(bv_to_int(A)).

gen(A,[]) => bv_eq(A,0).          % halt when A = 0
gen(A,[Z|Zs]) =>
    B = bv_take(A,3),             % B = A mod 8
    bv_xor(B,5,B1),               % B1 = B^5
    bv_pow(2,B1,P),               % P = 2**B1
    bv_div(A,P,C),                % C = A div P
    A1 = bv_drop(A,3),            % A1 = A/2**3
    bv_xor(B1,C,B2),              % B2 = B1^C
    bv_xor(B2,6,B3),              % B3 = B2^6
    Z #= 4*B3[3]+2*B3[2]+B3[1],   % out B3 mod 8
    gen(A1,Zs).
\end{verbatim}
\end{small}
\rule{\textwidth}{1pt}

\noindent
It encodes the CC program into a constraint network using bit vector and integer-domain constraints, and solves the constraints while minimizing the input. The \texttt{main} predicate creates a 56-bit bit-vector \texttt{A}, invokes \texttt{gen} to encode the sequence of instructions as constraints, calls \texttt{solve} to find a solution that minimizes \texttt{A}, and finally outputs the resulting value of \texttt{A}.

The \texttt{gen} predicate takes a bit vector representing the value in register \texttt{A} and a list of instructions from the given CC program. When the list is empty, the constraint \texttt{bv\_eq(A,0)} ensures that \texttt{A} is 0. Since the program generates a copy of itself, the sequence of instructions up to instruction \texttt{"out B3 mod 8"} executes once per code in the CC program.

Our Picat program took about 1 second to find the minimum input \texttt{109019476330651}. In terms of speed, our solution is not as fast as the DFS-based solution\footnote{https://www.reddit.com/r/adventofcode/comments/1hg38ah/comment/mb3vqjy/}, which only took 0.006 seconds, or the naive iterative solution \footnote{https://www.reddit.com/r/adventofcode/comments/1hg38ah/comment/m406q9f/}, which took 0.07 seconds. However, our solution is arguably more elegant than those solutions. Moreover, it could be more scalable than those solutions for large inputs.

\section{Day 24: Crossed Wires}
In Advent of Code 2024, Day 24, titled "Crossed Wires", the participants are tasked in the part one with modeling a Boolean circuit built from three types of logic gates (AND, OR and XOR). One example of such gate instance is given in Equation~\ref{eq:d24:example}. In this example, the wires named $abc$ and $x32$ are the inputs to the gate and the wire $def$ is the output of the gate.
\begin{equation}
\mbox{abc XOR x32} \longrightarrow \mbox{def}
\label{eq:d24:example}
\end{equation}
Part-1 requires modeling one single evaluation of a given circuit on a given input. This task can be easily tackled with any programming languages.

Part-2 specifies that the circuit is meant to be a binary addition circuit. More precisely, given the problem input, it should compute $z = x+y$, where $x = x_{44}..x_0$ is a binary number of 45 bits, where $x_0$ is the least significant bit, $y = y_{44}..y_0$ is another 45-bit number, and the result is $z=z_{44}..z_0$. However, the circuit is faulty, with the outputs of four pairs of gates swapped. The task is to identify which gates are involved in those swaps, and swap the outputs back so that the circuit functions correctly for every inputs.

The complexity of the problem arises from the presence of $n>300$ gates in the circuit. It is infeasible to solve the problem by brute force, as one would need to try the number of combinations given in Equation~\ref{eq:d24:comb}.
\begin{equation}
\frac{n!}{2^4\cdot 4!\cdot(n-8)!}>10^{17}\label{eq:d24:comb}
\end{equation}

This problem can be modeled as a constraint satisfaction problem. A straightforward model uses a binary variable for each pair of distinct gates, which indicates if the pair is involved in a swap. However, this model is inefficient as it requires 44000 binary variables.

Our model treats the problem as an assignment problem. It uses 8 slots, numbered 1 through 8, assuming slots $i$ and $i+1$ are involved in a swap, for $i \in \{1, 3, 5, 7\}$, and maps the slots to gates. To figure out the mapping, one can use one binary variable for each slot and each gate. In our Picat solution,\footnote{\url{https://github.com/cgrozea/AdventOfCode2024/blob/main/24/cleanup/part2article.pi}} presented below, an integer-domain variable is used for each slot, which maps the slot to a distinct gate.

\rule{\textwidth}{1pt}
\begin{small}
\begin{Verbatim}
% The core of the Picat program for Crossed Wires.

main([File])=>
    read_input(File,Gates),
    N = len(Gates),
    Slots = new_array(8),
    Slots :: 1..N,                            % each slot is mapped to a distinct gate
    all_different(Slots),                     
    NInputs = 40,                             % train the circuit using 40 random inputs
    foreach(_ in 1..NInputs)                    
        Vx := {random(0,1) : _ in 1..45},     % random input for X
        Vy := {random(0,1) : _ in 1..45},     % random input for Y
        Vz := new_array(45),
        Vz :: 0..1,                           % output Z
        add_gate_constrs(Gates,Slots,Vx,Vy,Vz),
    end,
    solve(Slots),
    output(Slots, Gates).

\end{Verbatim}
\end{small}
\rule{\textwidth}{1pt}

The program takes the input file name \texttt{File} as a command line argument.  It reads \texttt{Gates} from the file. It uses an array of 8 integer-domain variables, each of which has the domain \texttt{1..N}, where \texttt{N} is the number of gates. The \texttt{all\_different} constraint ensures that each slot is mapped to a distinct gate. 

The program generates a number of training inputs and calls \texttt{add\_gate\_constrs} to ensure that the circuit produces the expected output with the swaps. It is generally difficult to estimate in advance the number of training inputs needed. We need to run the program with a number of inputs that is large enough to produce a unique answer.

For each training input, a binary variable is created for each wire. The \texttt{add\_gate\_constrs} predicate converts the gates to constraints. For example, for the gate in \ref{eq:d24:example}, it generates the constraint
\begin{verbatim}
    Babc #^ Bx32 #= Bdef
\end{verbatim}
where the binary variable's name of each wire is the wire's name preceded with the letter \texttt{B}. For the very first layer of the circuit, the input wires are not involved in any swaps, so the values are taken from \texttt{Vx} and \texttt{Vy}. For any other wire of a gate,  we need to determine its binary variable based on whether or not the gate is involved in a swap. If the gate is not involved in any swaps, we use the variable of the wire itself. Otherwise, we need to use its partner wire's binary variable. The \texttt{add\_gate\_constrs} predicate also ensures that adding \texttt{Vx} and \texttt{Vy} produces \texttt{Vz}, which can be forced by using the bit-vector addition constraint \texttt{bv\_add(Vx,Vy,Vz)}.

The statistics shows that the Day 24 problem is the toughest AoC problem in 2024. Our model is encoded into SAT with $996000$ clauses and $361000$ variables. Despite the large encoding size, it is solved in 3.6 seconds. This is a testimony to the competitiveness of the Picat's SAT encoder and the Kissat solver. The straightforward model that uses a binary variable for each pair of gates fails to solve the problem in 2 hours.

Many of the approaches available on Reddit for Day 24 \footnote{\url{https://www.reddit.com/r/adventofcode/comments/1hl698z/2024_day_24_solutions/}} required manual inspection of the problem instance, and can't solve all instances that follow the problem description (see our instances generator\footnote{\url{https://github.com/cgrozea/AdventOfCode2024/blob/main/24/gen.pi}}). The program by \texttt{ayoubzulfiqar}\footnote{\url{https://github.com/ayoubzulfiqar/advent-of-code/tree/main/2024/Python/Day24}} for the problem is much faster than our SAT-based solution (it solves the problem in 0.06 seconds). It performs the ripple-carry algorithm on the inputs and identify the potential wires that could be swapped. While the solution is very intelligent and efficient, it requires genius insights and is limited by the assumption that the circuit follows a standard pattern repeated for every bit position.

\section{Day 16: Reindeer Maze}
In Advent of Code 2024, Day 16, participants are presented with a challenge titled "Reindeer Maze," which involves navigating a complex maze to guide a reindeer from a starting point to a destination. The problem is divided into two parts, each introducing unique objectives and constraints.

Part-1 requires participants to determine the minimum score needed for the reindeer to reach the destination. The maze is represented as a grid, which the reindeer can traverse in any of the four cardinal directions -- north, east, south and west -- as well as rotate left or right by 90 degrees. The objective is to find the path from the starting point to the destination that yields the lowest cumulative score. One slight complication is that orientation must be tracked to produce the correct result -- that is, if the reindeer is facing east and wants to move south, it must first turn clockwise, then move forwards. Turning has a significantly greater cost than moving forwards, so an optimal path minimizes the number of rotations. The following path-finding program using tabling can be easily adapted for this problem:
\rule{\textwidth}{1pt}
\begin{small}
\begin{verbatim}
    table(+,+,-,min)
    path(State,State,Path,Cost) =>
        Path = [],
        Cost = 0.
    path(State,Goal,Path,Cost) =>
        Path = [(State,NextState)|Path1],
        edge(State,NextState,ThisCost),
        path(NextState,Goal,Path1,Cost1),
        Cost = ThisCost+Cost1.
\end{verbatim}
\end{small}
\rule{\textwidth}{1pt}
The predicate \texttt{edge(State,NextState,Cost)} specifies a weighted directed graph, where \texttt{Cost} is the weight of the edge  from \texttt{State} to \texttt{NextState}. The predicate \texttt{path(State,Goal,Path,Cost)} finds a path \texttt{Path} from \texttt{State} to \texttt{Goal} with the minimum cost. The table mode denotes that the first two arguments are inputs, the third one is an output, and the last argument is to be minimized, meaning that for each input pair, the system stores only one path with the minimum cost. 

Part-2 extends the challenge by asking participants to identify all nodes that lie on \textit{any} of the shortest paths from the start to the destination. This requires not only finding the minimum score but also determining which nodes are part of paths that achieve this score.

An easy approach is to modify the above example, letting a state carry a via point. A node $N$ is on a shortest path if the shortest path cost from the start node to the destination with $N$ as a via point is globally minimum. However, this approach is inefficient as it significantly increases the number of states to be explored.

A strategic approach involves a combination of reverse traversals from the destination back to all intermediate nodes together with forward traversals from the start node as follows:

\begin{enumerate}
    \item \textbf{Forward traversals:} Determine the move costs $C_{SN}$ from the start node $S$ to each reachable node $N$ in the grid.
    \item \textbf{Backward traversals:} Determine the reverse move costs $C_{ND}$ from the destination node $D$ to each reachable node $N$ in the grid, in each of the four possible orientations (since it may be possible to arrive at the destination from multiple directions).
    \item \textbf{Count nodes on optimal paths:} Find all unique nodes that belong to an optimal path by searching the cost table for each node $N$ satisfying the equation $C_{SN} + C_{ND} = {MinCost}$, where ${MinCost}$ is the minimum path cost found in Part-1.
\end{enumerate}

The following shows an implementation of the approach using tabling:\footnote{The full source is available at: \url{https://github.com/nfzhou/aoc/blob/main/aoc_24_16_part2.pi}}

\noindent
\rule{\textwidth}{1pt}
\begin{small}
\begin{verbatim}
% A Picat program using tabling for Reindeer Maze.

main([File]) =>
    read_input(File,M,NRows,NCols,RStart,CStart,REnd,CEnd),
    Count = 2,
    foreach (R in 2..NRows-1, C in 2..NCols-1, 
             (R,C) != (RStart,CStart), (R,C) != (REnd,CEnd), M[R,C] !== '#')
        if minof(best_path_via(R,C,M,RStart,CStart,REnd,CEnd,Cost),Cost), Cost == 94444 then
            Count := Count+1,
        end
    end,
    println(Count).

% the best path from (RStart,CStart) to (REnd,CEnd) via (R,C)
best_path_via(R,C,M,RStart,CStart,REnd,CEnd,Cost) =>
    member([Dr,Dc], [[-1,0],[1,0],[0,-1],[0,1]]),
    best_path_to_start([Dr,Dc,R,C,M,RStart,CStart],Cost1),
    best_path_to_end([-Dr,-Dc,R,C,M,REnd,CEnd],Cost2),  % change direction by 180 degree
    Cost = Cost1+Cost2.
    
table (+,min)
best_path_to_start([Dr,Dc,R,C,M,R,C],Cost), Dr == 0, Dc == -1 =>  % direction must be west (0,-1)
    Cost = 0.  
best_path_to_start([Dr,Dc|T],Cost) ?=>                         % rotate
    turn(Dr,Dc,Dr1,Dc1),
    best_path_to_start([Dr1,Dc1|T],Cost1),
    Cost = Cost1+1000.
best_path_to_start([Dr,Dc,R,C,M|T],Cost) =>                    % move
    R1 = R+Dr,
    C1 = C+Dc,
    M[R1,C1] != '#',
    best_path_to_start([Dr,Dc,R1,C1,M|T],Cost1),
    Cost = Cost1+1.

table (+,min)
best_path_to_end([Dr,Dc,R,C,M,R,C],Cost) => Cost = 0.
best_path_to_end([Dr,Dc|T],Cost) ?=>                           % rotate
    turn(Dr,Dc,Dr1,Dc1),
    best_path_to_end([Dr1,Dc1|T],Cost1),
    Cost = Cost1+1000.
best_path_to_end([Dr,Dc,R,C,M|T],Cost) =>                      % move
    R1 = R+Dr,
    C1 = C+Dc,
    M[R1,C1] != '#',
    best_path_to_end([Dr,Dc,R1,C1,M|T],Cost1),
    Cost = Cost1+1.
\end{verbatim}
\end{small}
\rule{\textwidth}{1pt}

\noindent
The \texttt{read\_input(F,M,NRows,NCols,RStart,CStart,REnd,CEnd)} predicate reads the maze from \texttt{File} into matrix \texttt{M}, where \texttt{NRows} is the number of rows of \texttt{M}, \texttt{NCols} is the number of columns of \texttt{M}, \texttt{(RStart,CStart)} is the starting node, and \texttt{(REnd,CEnd)} is the destination.

The \texttt{main} predicate uses a loop to count the number of nodes on the shortest paths. It initializes \texttt{Count} to 2, counting in the starting and destination nodes. For each node \texttt{(R,C)} that is not the starting node, the destination, or an obstacle, if the cost of the best path from the starting node to the destination via node \texttt{(R,C)} equals the shortest-path's cost (94444, the result of Part-1), it increments \texttt{Count}.

The \texttt{best\_path\_via(R,C,M,RStart,CStart,REnd,CEnd,Cost)} predicate finds the best-path cost from \texttt{(RStart,CStart)} to \texttt{(REnd,CEnd)} via \texttt{(R,C)}. In order to facilitate sharing for tabled calls, it finds the best path cost from \texttt{(R,C)} to \texttt{(RStart,CStart)}, and the best path cost from \texttt{(R,C)} to \texttt{(REnd,CEnd)}. It uses deltas to represent directions. For example, \texttt{(Dr,Dc) = (0,-1)} represents the westward direction. In order to ensure that the forward and backward paths can be connected to form a viable path, it starts the two searches with opposite initial directions. The \texttt{best\_path\_to\_start} and \texttt{best\_path\_to\_end} are defined in the same way as the \texttt{path} predicate given above. The goal condition for \texttt{best\_path\_to\_start} also ensures that the direction is west, as the Reindeer faces east in the beginning.

As the \texttt{best\_path\_via} may return multiple shortest path costs due the nondeterministic choice for the starting direction for the backward search\footnote{The call \texttt{member([Dr,Dc], [[-1,0],[1,0],[0,-1],[0,1]])} is nondeterministic.}, the program uses \texttt{minof} to retrieve the minimum of such path costs.

The implementation using tabling took 16 seconds to find the count 502 for the input maze. We also implemented the approach in Picat using Dijkstra's algorithm\footnote{\url{https://github.com/DestyNova/advent_of_code_2024/blob/main/16/part2_dijkstra.pi}}. This implementation only took 1 second to find the answer.

%This analysis ensures that all minimal paths are accounted for, providing a complete solution to the problem. The source code\footnote{\url{https://github.com/DestyNova/advent_of_code_2024/blob/main/16/part2_dijkstra.pi}} for this implementation is quite concise and relies on a \texttt{planner_star} module which mimics Picat's built-in \texttt{planner} API but uses Dijkstra and A* search instead of Prolog-style backtracking to select optimal moves. In this case we don't use the planner's \texttt{action} predicate, but instead specify \texttt{successor} and \texttt{predecessor} predicates which are passed to the Dijkstra search routine.

\section{Day 21: Keypad Conundrum}
In Advent of Code 2024, Day 21, titled "Keypad Conundrum," participants are tasked with navigating a series of interconnected keypads to input specific codes, with each part of the challenge introducing additional layers of complexity.

In Part-1, the initial scenario involves a directional keypad that the human controller uses (Figure \ref{fig:keypads}). The objective is to determine the shortest sequence of button presses required to instruct a robot to type a given code, such as '029A', on a numeric keypad (Figure \ref{fig:keypads}). The robot starts with its arm positioned over the 'A' button on the numeric pad. Each directional button moves the robot's arm in a specific direction and the activate button (A) instructs the robot to press the current button. The challenge is to compute the minimal sequence that results in the robot correctly inputting the desired code on the numeric keypad. Part-1 can be solved with brute-force search.

In Part-2, the complexity increases as additional layers are introduced. Now, a second robot is required to control the first robot, a third robot is required to control the second robot, so on up to the 26th robot, and human controller instructs the 26th robot. The task is to determine the shortest sequence of button presses for the human controller that will propagate through the chain of robots, ultimately resulting in the correct code being entered on the numeric keypad. 

The output is the sum of the complexities of the codes, where each code's complexity is equal to the result of multiplying the length of the shortest sequence for the human controller and the numeric part of the code.

\begin{figure}[t]
\begin{center}
\begin{verbatim}
                    +---+---+---+        +---+---+
                    | 7 | 8 | 9 |        | ^ | A |
                    +---+---+---+        +---+---+---+
                    | 4 | 5 | 6 |        | < | v | > |
                    +---+---+---+        +---+---+---+
                    | 1 | 2 | 3 |
                    +---+---+---+
                    |   | 0 | A |
                    +---+---+---+
\end{verbatim}
\caption{\label{fig:keypads}The numeric (left) and directional (right) keypads.}
\end{center}
\end{figure}

The following shows the main part of a Picat program for this problem.\footnote{The full source is available at: \url{https://github.com/nfzhou/aoc/blob/main/aoc_24_21_part2.pi}}

\noindent
\rule{\textwidth}{1pt}
\begin{small}
\begin{verbatim}
% A Picat program using tabling for Keypad Conundrum.

main([File]) =>
    Codes = read_file_lines(File),
    Comp = 0,
    foreach (Code in Codes)
        num_pad_plan(t,'A',Code,Plan,_),
        trans_plan_rec(25,Plan,Len),
        Comp := Comp + to_int([D : D in Code, digit(D)]) * Len
    end,
    println(Comp).

table (+,+,+,-,min)
num_pad_plan(PreDir,_,[],Plan,Obj) => 
    Plan = [],
    Obj = (0, 0).
num_pad_plan(PreDir,C,[C|Code],Plan,Obj) =>
    Plan = ['A'|Plan1],
    num_pad_plan(t,C,Code,Plan1,Obj1),
    Obj1 = (Len1, Turn1),
    Obj = (Len1+1,Turn1).
num_pad_plan(PreDir,C,Code,Plan,Obj) =>
    member(Dir,[l,u,d,r]),   % best ordering
    move_num_pad(Dir,C,NextC),
    Plan = [Dir|Plan1],
    num_pad_plan(Dir,NextC,Code,Plan1,Obj1),
    Obj1 = (Len1, Turn1),
    Obj = (Len1+1,cond(Dir==PreDir,Turn1,Turn1+1)).

table
trans_plan_rec(0,Plan,Len) =>
    Len = len(Plan).    
trans_plan_rec(N,[],Len) => Len = 0.
trans_plan_rec(N,Plan,Len) =>
    extract_chunk(Plan,PlanR,Chunk),
    dir_pad_plan(t,'A',Chunk,ChunkPlan,_),
    trans_plan_rec(N-1,ChunkPlan,Len1),
    trans_plan_rec(N,PlanR,Len2),
    Len is Len1+Len2.
\end{verbatim}
\end{small}
\rule{\textwidth}{1pt}

\noindent
The \texttt{main} predicate reads codes from the input file and iterates over them using a \texttt{foreach} loop. For each code, the predicate \texttt{num\_pad\_plan} computes an optimal plan to instruct the numeric pad robot to generate the code. This plan is then recursively translated through 25 control layers by the predicate \texttt{trans\_plan\_rec}, producing a final plan for the human controller. An assignment statement computes the code’s complexity and adds it to the variable \texttt{comp}, which accumulates the total complexity.

The predicate \texttt{num\_pad\_plan(PreDir,C,Code,Plan,Obj)} takes three inputs: \texttt{PreDir}, the previous direction of movement of the numeric robot; \texttt{C}, the robot’s current key position; and \texttt{Code}, the sequence to be typed. It produces a plan \texttt{Plan} that minimizes the objective value \texttt{Obj}, which is a pair \texttt{(Length, Turns)} representing the plan’s length and number of turns. The table mode declaration \texttt{(+,+,+,-,min)} specifies that the first three arguments are inputs, the last two are outputs, and \texttt{Obj} is to be minimized. Initially, the robot starts at key ‘A’ with a dummy direction \texttt{t} passed to \texttt{PreDir}. The definition of \texttt{num\_pad\_plan} is straightforward:

The first rule handles the base case where the code is empty. It returns an empty plan and the objective value \texttt{(0, 0)}.

The second rule applies when the robot is already at the correct key \texttt{C}. It adds ‘A’ to the plan to simulate pressing the key, and recursively calls itself to find an optimal plan \texttt{Plan1} and objective \texttt{Obj1} for the remaining code. The resulting \texttt{Obj} is derived by incrementing the plan length while preserving the number of turns.

The third rule handles the case where the robot is not on the required key. It nondeterministically chooses a direction in the order left (l), up (u), down (d), and right (r), then recursively computes a subplan \texttt{Plan1} and objective \texttt{Obj1} after the robot moves in that direction. The final objective \texttt{Obj} is calculated by incrementing both the length and the number of turns.

The predicate \texttt{trans\_plan\_rec(N, Plan, Len)} translates a plan from the numeric robot up the control chain to the \texttt{N}th controller:

The base case, when \texttt{N = 0}, terminates the recursion.

The general case processes the plan in chunks, where a chunk is the longest subplan that begins with a non-'A' symbol, terminates with 'A' symbols, and contains no intermediate 'A' symbols. The predicate \texttt{extract\_chunk(Plan, PlanR, Chunk)} splits \texttt{Plan} into \texttt{Chunk}, the first chunk, and \texttt{PlanR}, the remaining sequence. The predicate \texttt{dir\_pad\_plan(t, ‘A’, Chunk, ChunkPlan, \_)}—defined similarly to \texttt{num\_pad\_plan}—finds an optimal plan \texttt{ChunkPlan} for the next controller up the chain. The rule then recursively calls \texttt{trans\_plan\_rec} on \texttt{ChunkPlan} and \texttt{PlanR}, decrementing the level since \texttt{ChunkPlan} belongs to the upper controller.

The program computes the result 226179529377982 in just 0.037 seconds for the input list \texttt{["671A", "826A", "670A", "085A", "283A"]}. The solution leverages two key ideas. First, it limits direction choices to a fixed order [l, u, d, r] instead of trying all permutations, significantly reducing the search space. Second, it performs chunk-wise translation along the control chain, with optimal subplan contributing to an optimal overall solution. Additionally, chunking facilitates sharing of tabled calls, improving efficiency.

The Reddit's AoC 24 Day 21 Megathread\footnote{\url{https://www.reddit.com/r/adventofcode/comments/1hj2odw/2024_day_21_solutions/}} gives several solutions in other languages. Many approaches use dynamic programming to find a minimal path. The Python program by \texttt{YOM2\_UB}\footnote{\url{https://pastebin.com/rExTBYr6}} uses depth-first search together with memoization. It only took 0.031 seconds to solve the problem. The Rust program by \texttt{michelkraemer},\footnote{\url{https://github.com/michel-kraemer/adventofcode-rust/blob/main/2024/day21/src/main.rs}}, which also uses depth-first search with memoization, only took 0.013 seconds. While our Picat solution is not competitive in terms of speed, it is certainly more readable than those solutions, thanks to the built-in dynamic programming feature.

\section{Conclusion}
This paper presents Picat solutions for five problems selected from the 2024 Advent of Code collection. The solution to the Maximum Clique problem is primarily pedagogical, illustrating how to model and solve a simple combinatorial problem using SAT. The solution to the Chronospatial Computer reverse engineering problem translates a program into a constraint network using bit-vector constraints, allowing the inference of inputs from known outputs. The solution to the Crossed Wires problem showcases not only the power of SAT encoding and solving but also the critical role of effective modeling. The idea of using eight integer-domain variables to identify the swapped wires proves to be particularly effective.

The solution to the Reindeer Maze problem demonstrates the strength of a strategic reformulation: converting a one-to-many shortest paths problem into a many-to-one problem. By reversing the graph, the approach greatly enhances sharing among tabled calls. The Keypad Conundrum solution illustrates tabled search involving the simultaneous optimization of two objectives.

While the Picat solutions may not always be the shortest or fastest compared to those found online, they are declarative, readable, and do not rely on deep insights or clever tricks.

Many of the selected problems have real-world relevance. For instance, the reverse engineering problem relates to program analysis and verification, while the Crossed Wires problem mirrors fault diagnosis and troubleshooting tasks. These examples show that Picat can be a viable tool for such applications, much like other formal methods tools. Path-finding and planning problems are ubiquitous in practice. The solutions to the Reindeer Maze and Keypad Conundrum problems demonstrate how declarative dynamic programming can be expressed naturally in Picat, without requiring the implementation of complex algorithms or heuristics.
\nocite{*}
\bibliographystyle{eptcs}
\bibliography{generic}
\end{document}